\documentclass[conference]{IEEEtran}
\IEEEoverridecommandlockouts
\usepackage{mathtools}
\usepackage[utf8]{inputenc}
\usepackage{fancyhdr} 
\usepackage{balance}
\usepackage{cite}
\usepackage{amsmath,amssymb,amsfonts}
\usepackage{algorithmic}
\usepackage{graphicx}
\graphicspath{{image/}}
\usepackage{textcomp}
\usepackage{xcolor}
\usepackage[english]{babel}
\usepackage{multirow}
\usepackage{multicol}

\usepackage{tikz}
\usetikzlibrary{positioning}

\def\BibTeX{{\rm B\kern-.05em{\sc i\kern-.025em b}\kern-.08em
    T\kern-.1667em\lower.7ex\hbox{E}\kern-.125emX}}
\begin{document}

\title{Multi-criteria Rating and Review based Recommendation Model}

\author{
Emrul Hasan\\
 Department of Computer Science\\
 Toronto Metropolitan University\\
 Toronto, Canada\\
  \texttt{e1hasan@ryerson.ca}
  \and
Chen Ding\\
 Department of Computer Science\\
 Toronto Metropolitan University\\
 Toronto, Canada\\
  \texttt{cding@ryerson.ca}
  \and
Alfredo Cuzzocrea\\
 iDEA Lab\\
 University of Calabria\\
 Rende, Italy\\
    \texttt{alfredo.cuzzocrea@unical.it}
}

\maketitle

\begingroup\renewcommand\thefootnote{}\footnotetext{\copyright~2022 IEEE. Personal use of this material is permitted. Permission from IEEE must be obtained for all other uses, in any current or future media, including reprinting/republishing this material for advertising or promotional purposes, creating new collective works, for resale or redistribution to servers or lists, or reuse of any copyrighted component of this work in other works. This is the accepted version of the paper published in the 2022 IEEE International Conference on Big Data (Big Data), pp. 5494--5503, doi: 10.1109/BigData55660.2022.10020741.}\addtocounter{footnote}{-1}\endgroup

\textbf{Abstract}--
These days, due to the advancement of information technology, recommendation system has become one of the key tools for e-commerce business. E-commerce platforms allow users to provide feedback in both written comments and numerical ratings. Recommendation systems are utilized to recommend users to new or unseen items based on these previously collected comments or ratings.
In recent years, multi-aspect or multi-criteria based recommendation systems have been studied a lot by the recommendation research community. However, these research works are conducted either with reviews or ratings, not with both.
In this project, we argue that integrating both textual reviews (with multiple aspects) and numerical multi-criteria ratings can further enhance the overall rating prediction accuracy. We propose a Multi-criteria Rating and Review based Recommendation model (MRRRec). We show that incorporating multi-criteria ratings into multi-aspect ratings from reviews has a great impact on performance. Our proposed model outperforms several state-of-the-art models such as ANR, DeepCoNN, and Deep Multi-criteria Recommendation System in terms of MSE, MAE, precision, recall, and F1. We show that our proposed model achieves an average of 19\% and 23.0\% lower MSE and MAE respectively and 7.0\%, 1.0\%  and 3.8\% higher precision, recall, and F1 score respectively. We further show that our model performs significantly better with Word2Vec word embedding than the GloVe word embedding method.


\section{Introduction}
The recommendation system provides a personalized recommendations to users based on historical data. They are broadly classified as collaborative filtering, content-based filtering, knowledge-based filtering and hybrid recommendation  system\cite{principles, kim2019}. Collaborative Filtering (CF) techniques\cite{raghuwanshi2019collaborative} are the most widely used methods. CF uses the user's past historical data to make recommendations. It can be further classified into two categories, user-based collaborative filtering, and item-based collaborative filtering. User-based collaborative filtering uses user similarities whereas item-based filtering uses item similarities to make a recommendation. Traditionally the most popular CF techniques are based on Matrix factorization (MF) \cite{Koren2009}, which decomposes the user-item rating matrix into two smaller matrices. Although MF is a widely used recommendation technique, because of its dependency on the user's explicit ratings, MF suffers from data sparsity problem \cite{grvcar2005data} as, in real-world scenarios, the user-item rating matrix is often a sparse matrix. Furthermore, a single rating or binary interaction with an item also cannot capture the user's fine-grained preferences on various aspects of the item.

Different data mining techniques \cite{cuzzocrea2013analytics, ceci2015effectively,cuzzocrea2006accuracy} can be applied to build recommendation models from a large dataset. To deal with data sparsity problems, various approaches have been attempted. For example incorporating textual reviews\cite{zhang2016integrating}, images\cite{cheng2019mmalfm}, and contextual information\cite{cheng2014just} into MF can provide recommendations with higher accuracy, scalability, and explainability \cite{chen2015recommender,cheng2019mmalfm, he2015trirank }. 
Recently, aspect-based \cite{guan2019attentive, chin2018anr, li2020aspect} and multi-criteria rating \cite{nassar2020multi, hamada2018artificial,hong2021clustptf,shambour2021deep,gupta2020credibility,hong2021multi,9005677} based recommendation systems have shown promising performance in mitigating the data sparsity and cold start problems. 

The tourism management platform, TripAdvisor, collects feedback from customers in three different modes including overall ratings, textual comments, and criteria ratings. An example of a hotel review is shown in Figure 1 where a customer provided feedback on three different categories, overall rating (5 stars) at the top, textual comment in the middle, and ratings on individual criteria (5 stars to value, service, and cleanliness). We define the criteria ratings as explicit ratings because user provided ratings explicitly on individual criteria. Criteria ratings can also be extracted from textual comments. Aspects are considered as the terms or phrases or topics in the review that users are concerned about. Considering hotel domain, price, location, value, room size, service, etc. are considered as aspects. Each of the aspects consists of a set of one or more aspect terms. For example, aspect price encompasses a set of aspect terms: $\{expensive, cheap, reasonable, \dots\}$. Associated rating with these aspects can also be estimated \cite{chin2018anr, wu2019arp}. We define these aspects ratings as implicit criteria ratings because these ratings are not provided by users explicitly. It is important to note that these estimated implicit criteria may or may not be similar to the explicit criteria ratings. 

Multi-criteria rating-based recommendation system uses criteria ratings to predict the overall rating. In this method, multi-criteria ratings are used to predict the overall ratings. Although these methods have achieved success in solving the cold start problem, they ignore users' fine-grained opinions expressed through textual comments. On the other hand, aspect-based recommendation systems utilize users' review texts to predict overall ratings. Though aspect based method performs well in predicting overall rating, these methods ignore the users' opinions expressed through explicit criteria ratings, resulting poor accuracy measure in overall rating prediction.

\begin{figure}[ht]
\centering
\includegraphics[width=0.45\textwidth]{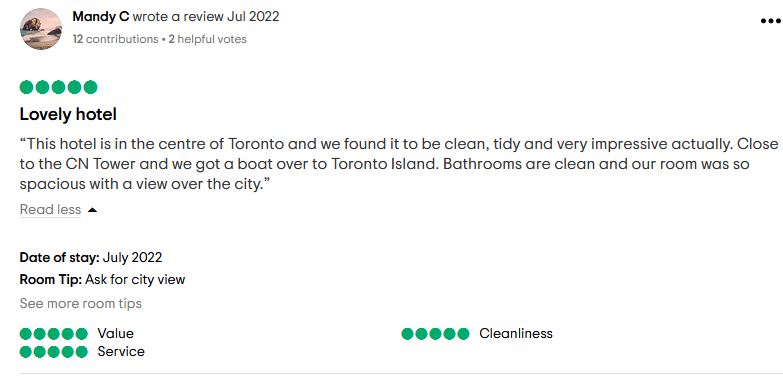}
    \caption{Customer review on hotel}
\end{figure}

We argue that the integration of implicit criteria ratings into explicit multi-criteria ratings can improve the performance on overall rating prediction. Real-world publicly available Tripadvisor dataset \cite{tripAdvisor}, provides users' feedback as textual comments, multi-criteria ratings, and overall ratings. In this paper, we propose a novel solution for overall rating prediction from explicit and implicit multi-criteria ratings. Our model consists of two parallel paths for computing implicit and explicit criteria rating, and finally a DNN (Deep Neural Network) for computing overall rating. 

The main contributions of our work are listed below:

\begin{enumerate}
   \item  We have proposed a novel Multi-criteria Rating and Review based Recommendation model (MRRRec), which computes implicit ratings and explicit criteria ratings leading to the overall rating prediction. We compute implicit criteria rating by the aspect-level user and item representation, and user and item importance estimation \cite{chin2018anr}. To the best of our knowledge, this is the first paper that computes overall ratings by estimating both implicit and explicit criteria ratings. 
   \item Our proposed MRRRec model is experimented with a publicly available Tripadvisor dataset and compared performance with several state-of-the-art baseline methods, namely, DeepCoNN, ANR, and Deep Multi-criteria RS. We show that our model outperforms all the state-of-the-art methods in terms of MSE, MAE, precision, recall, and F1 measure.
   
   \item We further compared the performance of our model using two different word embedding methods, Word2Vec and GloVe word embedding. It is observed that the best performance is achieved with Word2Vec word embedding method. To achieve the best performance, we also experimented model with different combinations of hidden layers in the DNN structure.

\end{enumerate}

The rest of the paper is organized as follows. In section II, we discuss the related work. Section III is devoted to the detailed description of the model. Extensive experiments and results are demonstrated in section IV and V respectively. Section VI presents the conclusion and future works.

\section{Related Work}
In this section, we briefly explain the areas that are highly related to our proposed work, aspect-based recommendation system, multi-criteria rating recommendation system, and deep learning-based recommendation. 

\subsection{Aspect-based Recommendation System}
Aspect-based methods retrieve the aspects from the customers' textual reviews and leverage them for rating prediction. Generally, two methods are used for aspect-based recommendation systems. The first one is extracting the aspects from the textual review explicitly\cite{zhang2014explicit} and the second one is the aspect representation learning\cite{chin2018anr}.
Aspect extraction methods often use external sentiment analysis tools\cite{tabassum2016opinion} to retrieve the latent aspect. These methods are highly dependent on the quality of the sentiment analysis tools and the application domain. On the other hand, aspect learning methods learn the aspect from the textual reviews. Aminu Da'u et al. \cite{da2020weighted} used multi-channel CNN model for aspect-based opinion mining and finally an aspect weighted rating estimation method was developed using tensor factorization techniques. Wu et al.\cite{wu2019arp}
proposed an aspect-aware neural review rating prediction model where review-level rating predictors and multiple aspect-level rating predictors are jointly trained by a collaborative learning framework. Cheng et al. \cite{cheng2018aspect} proposed an aspect-aware topic model on review text which estimates the aspect importance of a user towards an item. It does the automatic aspect extraction and aspect importance estimation for each user-item pair which can solve the data sparsity problem in matrix factorization models. Li et al. \cite{li2019latent} proposed a latent multi-criteria recommendation model. In this method, latent multiple criteria were captured from user reviews using a variational auto-encoder. The model automatically generates multi-criteria ratings by combining autoencoding and embedding compression techniques.

\subsection{Multi-criteria Rating-based Recommendation System}
Multi-criteria rating-based recommendation system focuses on utilizing criteria ratings for predicting overall ratings as well as making recommendations.
Nassar et al. \cite{nassar2020multi} used a deep neural network to predict the overall rating from criteria ratings. In this method, overall ratings are predicted with a two-step process. In the first step, a pure collaborative filtering method is applied to predict the unknown criteria ratings where user and item IDs are considered as features and used as input to the Deep Neural Network. In the second part, those predicted criteria ratings are used as input to the overall rating prediction DNN model. Hong et al.\cite{hong2021multi} proposed a multi-criteria tensor model for tourism recommender systems. A Cluster-based Parallel Tensor Factorization (ClustPTF) was proposed by Hong et. al. \cite{hong2021clustptf}. ClustPTF utilized sentiment analysis technique to alleviate the data sparsity in the tensor model. K-means clustering was used for grouping similar user preferences which lead to improvement in the recommendation diversity.

Liu et al. \cite{liu2019daml} proposed DAML, a Dual Attention Mutual Leaning model which unifies ratings and reviews for a recommendation system. In this model, multi-criteria ratings are incorporated into the user-item rating prediction tensor model. A credibility score-based multi-criteria recommendation system was proposed by Gupta et al. \cite{gupta2020credibility}. This method focuses on the credibility score of an item based on multi-criteria ratings. Personal and public preferences are considered to compute the credibility score. Shambour \cite{shambour2021deep} developed a deep learning-based multi-criteria recommendation system that employs a deep auto-encoder to extract the user's latent preferences on an item. An Artificial Neural Network and particle Swarm optimization algorithm-based multi-criteria recommendation system was proposed by Hamada et al.\cite{hamada2018artificial}. In this method, the n-dimensional multi-criteria rating problem is decomposed into a single rating problem and the prediction function was built using a neural network.

\subsection{Deep Learning-based Recommendation System}
Historically, matrix factorization based standard machine learning models were popular for interaction-based recommendation system \cite{mnih2007probabilistic, he2016fast}. Due to the high dependency on users' explicit ratings, MF suffers from data sparsity problems. These days deep Learning has achieved great success in many areas of computer science and artificial intelligence, including but not limited to Natural Language Processing \cite{kulkarni2021deep,alcamo2021sentiment}, Computer Vision\cite{wu2017application}, Question Answering \cite{lahiri2022gameofthronesqa, raza2022coquad}, and Speech Recognition\cite{kamath2019deep}. Recently, deep learning has proven to achieve state-of-the-art performance in recommendation systems \cite{9378340,seo2017interpretable,zheng2017joint,8596492,9005633,9671467}. Various deep-learning models are applied to extract latent features from textual reviews. For example, Bugao et al. \cite{li2021aspect} employed Convolutional Neural Network(CNN) and a Long Short Term Memory (LSTM) to reveal the local and global aspects respectively. CNN and LSTM are applied in two parallel paths to capture the hidden local and global topics in the reviews. In the final step, local and global aspect features are merged through mutual interactions for rating prediction. In \cite{seo2017interpretable}, Seo et. al. proposed a review rating prediction model that models user preferences and item properties with the help of location and global attention mechanisms. This method can extract complex features from review texts. Local attention captures the users' latent preferences and items' latent characteristics, and global attention focuses on the semantic meaning of the whole review text. The neural attention mechanism is proven to be a great tool for explainable and better-performing recommendations. In \cite{chen2018neural}, the neural attention mechanism is used to identify the usefulness of the review. NRPA \cite{liu2019nrpa} is another neural recommendation model where personalized attention is used to learn the user and item representation from reviews. The key focus of the neural attention model is to urge on segmenting documents by weighting each word by a learned attention vector. 
For our model, we consider both explicit (multi-criteria ratings) and implicit (aspect-based ratings) rating-based ratings as well as overall rating prediction using a deep neural network.



\section{Proposed Model - MRRRec}
In this section, we present our proposed Multi-criteria Review and Rating based Recommendation model (MRRRec). Before elaborating the model detail, first we define the problem statement and then describe the different components of the model. Next, we present related theories. Finally, we explain the effect of different parameters.

\subsection{Problem Definition}
Let us consider a corpus $D$ with a set of reviews 
$X: \{x_1, x_2, x_3,\dots\,x_n\}$, a set of explicit criteria ratings $C: \{c_1^J, c_2^J, \dots\,c_n^J\}$ where $J: \{j_1, j_2, j_3, \dots\,j_{|J|}\}$ are a set of explicit criteria for each review, and a set of overall ratings $R:\{r_1, r_2, r_3, \dots\,r_n\}$ given by a set of users $U: \{u_1, u_2, u_3,\dots\,u_{|U|}\}$ to a set of items $I: \{i_1, i_2, i_3,\dots\,i_{|I|}\}$. In our case, hotels are the items. We assume that each review contains a set of aspects $K: \{k_1, k_2,k_3,\dots\,k_{|K|}\}$. Therefore, a set of implicit criteria ratings can be defined as  $A: \{a_1^K, a_2^K, a_3^K,\dots\,a_n^K\}$. It is important to note that $J$ and $K$ may or may not be the same. Table I summarizes the key notations used throughout this paper.

\begin{figure*}
\centering
\includegraphics[width=1.0\textwidth]{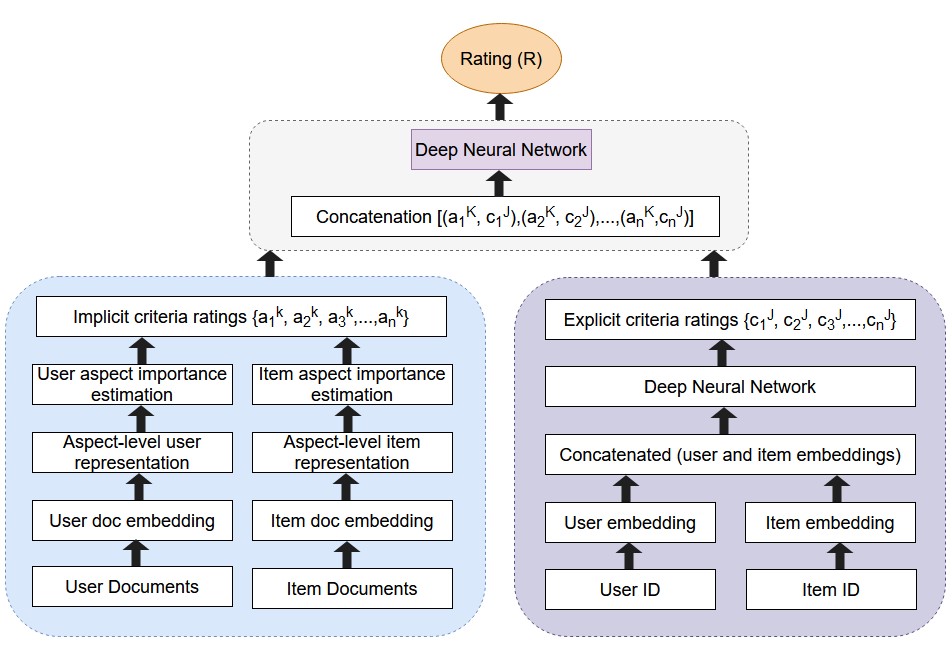}
    \caption{Architecture of the proposed MRRRec model}
\end{figure*}

\begin{table}
\centering
\caption{Definition of various notations}
\begin{tabular}{c c }
Notations &Definition\\
\hline
$D$& corpus with criteria ratings, textual review and overall ratings.\\
$X$ &set of reviews\\
$C$ & set of criteria ratings\\
$n$ & n is the number of reviews\\
$|J|$  & number of explicit criteria ratings for each review\\
$R$  & a set of overall ratings \\
$U$  & a set of users\\
$|U|$  & number of users\\
$I$  & a set of items\\
$|I|$  & number of items\\
$|K|$ & number of implicit criteria for each review\\
$D_{U}$ & user documents (set of reviews from users $U$)\\
$D_{I}$ & item documents (set of reviews for items $I$)\\
\hline
$P_{U,K}$  & aspect-level user representation\\
$Q_{I,K}$  & aspect-level item representation\\
$\beta_{U,K}$  & user aspect importance\\
$\beta_{I,K}$  & item aspect importance\\
$U_{\mu}$& user id embedding matrix\\
$I_{\mu}$& item id embedding matrix\\
$W_{l}$& weight matrix of hidden layer $l$\\
$h$& hidden layer\\
$L$& number of hidden layers\\
$b_{l}$& bias for hidden layer $l$\\
$\chi$ & concatenated user and item id embedding matrix\\
$Z$& input features for overall rating prediction DNN\\
\hline
\end{tabular}
\end{table}

Our ultimate goal is to predict overall rating $\hat{R}$ for user-item pairs $(U,I)$. The end goal can be divided into three sub tasks as described below. 

\begin{enumerate}
    \item Explicit criteria rating prediction: we predict the unknown explicit criteria ratings, $C$, using a deep neural network. In this case, user and item IDs are considered as features for DNN.
    \item Implicit criteria rating prediction: implicit criteria ratings, $A$, are estimated with aspect-based representation learning for both users and items via an attention-based component. In this case, user document $D_{U}$ and item document $D_{I}$ are used as features.
    \item Overall rating prediction: overall ratings, $\hat{R}$, is predicted from implicit and explicit criteria ratings using deep neural network.
\end{enumerate}

\subsection{Architecture}
The detailed architecture of our proposed MRRRec model is shown in Figure 2. The model consists of two parallel paths and a deep neural network at the top. Path one, the left side of the network, predicts the implicit ratings from the review, and path two, the right side of the network, predicts explicit criteria ratings. The top part of the model learns the overall rating via a deep neural network.

\subsection{Implicit Criteria Rating Prediction}
To predict the implicit criteria ratings, we follow the work of Chin et al. \cite{chin2018anr}. In this method, user and item documents are used as features of the model. User document $D_{U}$ represents a set of reviews written by a set of users $U$ and item document $D_{I}$ is a set of reviews for a set of items $I$. First of all, user and item documents are transformed to embedding vectors using pre-trained word embedding GloVe\cite{pennington2014glove} or Word2Vec\cite{church2017word2vec} model. In the next step, the neural attention mechanism is applied to learn the aspect-level user and item representation followed by aspect importance estimation. For a set of users, $U$ and a set of items $I$, a set of implicit criteria ratings $A$ is defined as

\begin{eqnarray}
A=\beta_{U,K} \cdot \beta_{I,K} \cdot (P_{U, K}(Q_{I,K})^\top)
\end{eqnarray}

where $\beta_{U,K}$ represents a set of user aspect importance or a set of users $U's$ satisfaction towards a set of $K$ aspects and $\beta_{I,K}$ represents the importance of a set of $K$ aspects for a set of items $I$. $P_{U, K}$ and $Q_{I,K}$ are defined as the aspect-level user and ite the representations respectively. Detailed calculation can be found in paper \cite{chin2018anr}.

\subsection{Explicit Criteria Rating Prediction}
In this case, we compute a set of explicit criteria ratings $C$ for a set of users $U$ on a set of items $I$. This part of the model is based on the 
paper by Nassar et al.\cite{nassar2020novel}. User IDs and item IDs are used as input features. According to He et al. \cite{he2017neural} categorical features such as IDs that do not have logical order can be represented as features. We embed the IDs and represent them as low-dimensional vectors. Embedding vectors are initialized with random initialization and values are adjusted during training steps. User embedding and item embedding are concatenated before feeding them into neural networks. Concatenated embedded feature vectors are then passed through a series of hidden layers. Consider the user ID vector $U_{\mu}$ and item ID vector $I_{\mu}$. These two vectors are concatenated as follows.
\begin{eqnarray}
\chi =Concatenate(U_{\mu}, I_{\mu})
\end{eqnarray}
where $U_{\mu}$ represents the user ID embedding matrix and $I_{\mu}$ represents the item ID embedding matrix.

The feature vector $\chi$ is the input to the DNN. ReLU (Rectified Linear Unit)\cite{nair2010rectified} is used as the activation function. ReLU is considered the most efficient activation function.
\begin{eqnarray}
ReLU (z)=max(0,z)
\end{eqnarray}

Mathematically the output of the hidden layer is represented as 
\begin{eqnarray}
h_{l}=ReLU(W_{l}h_{l-1}+b_{l})
\end{eqnarray}

where $W_{l}$, $b_{l}$ and $h_{l}$ represent weights, bias and hidden layers respectively.
The final output of the explicit criteria ratings can be represented as follows: 
\begin{eqnarray}
C=ReLU(W_{L}h_{L-1}+b_{L})
\end{eqnarray}

where L is the number of layers. $W_{L}$ is the weight matrix of the final layer $L$.

\subsection{Overall Rating Prediction}
\begin{figure}[ht]
\centering
\includegraphics[width=0.45\textwidth]{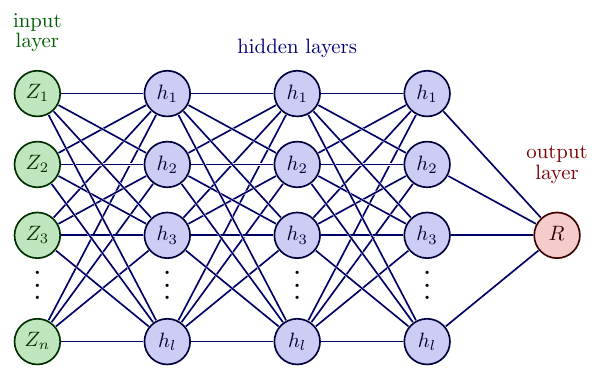}
    \caption{The detailed architecture of overall rating prediction DNN}
\end{figure}
The architecture of the overall rating prediction DNN model is shown in Figure 3. First of all, the explicit criteria ratings $C$ and implicit criteria ratings $A$ are concatenated. We normalize the concatenated ratings because DNN is sensitive to input distribution and scaling \cite{ioffe2015batch}.
\begin{eqnarray}
Z=Concatenate (C,A)
\end{eqnarray}
where $Z$ is the features for the overall rating prediction DNN. The dimension of the input layer in the DNN is the dimension of $Z$. The output dimension is 1 which is the overall rating. The output of the hidden layers is computed using equation 4. The overall ratings are predicted using equation 5 and is expressed as

\begin{eqnarray}
\hat{R}=ReLU(W_{L}h_{L-1}+b_{L})
\end{eqnarray}

where $\hat{R}$ is the final predicted rating and the input layer is $Z=h_{o}$.

\section{Experiments}

We evaluate the performance of our proposed model using the tripadvisor dataset \cite{tripAdvisor}. Tripadvisor is a tourism management online platform that offers hotel, flight, restaurant, cruise, and car rental facilities. The data was collected in an 8-year span, from 2004 to 2012. The dataset contains overall ratings, multi-criteria ratings, and textual reviews. There are a total of 8 different criteria and associated ratings including value, rooms, location, cleanliness, check-in/front desk, service, sleep quality, and business service. The ratings for individual criteria ratings and overall ratings range from -1 to 5. We remove the ratings below 1 and criteria with more than 75\% missing values. After removing missing data, we have 6,26,263 reviews and 5 criteria. For data processing, we remove punctuation and next-line characters. All reviews are tokenized using the natural language processing toolkit, NLTK\cite{loper2002nltk} and we build a vocabulary with the most frequent 50,000 words. We also remove the reviews of less than 10 tokens. For both user and item documents, each document length is truncated 500 tokens to keep consistent with ANR model\cite{chin2018anr}. The detailed statistics of the data are shown in Table II and the distribution of criteria ratings and overall ratings are shown in Table III. 

\begin{table}
\centering
\caption{Data statistics}
\begin{tabular}{c c }
\hline
Field description & Value\\
\hline
Number of users & 299,855\\
Number of items & 9,763\\
Overall ratings & 577,854\\
Density & 0.0001974\\
\hline
\end{tabular}
\end{table}

\begin{table}
\centering
\caption{Rating distribution}
\begin{tabular}{c c c c c c }
Criteria ratings &1 & 2 & 3 & 4 & 5\\
\hline
Service &31211& 37208&77336 & 154537&325971\\
Rooms & 32111 &  36629& 102368& 188780 & 266375\\
Location  & 10043 & 16267 & 63425 & 155110 & 381418\\
Value  & 35510 & 37996& 98085 & 185974 & 268698\\
Cleanliness   & 22441 & 22668 & 61721 & 157046 & 362387\\
Overall & 30699&34631&85853&201190&273890\\
\hline
\end{tabular}
\end{table}

\subsection{Baselines}
We compare the performance of our proposed MRRRec with the following baselines.
\begin{enumerate}
    \item \textbf{ANR}\cite{chin2018anr}. An aspect-based neural recommender which models the user and item documents for rating prediction. User and item documents are utilized to learn the aspect based representation followed by aspect importance estimation using attention-based component, and finally rating prediction.
    
    \item \textbf{DeepCoNN}\cite{he2017neural}. Deep Cooperative Neural Network (DeepCoNN) is one of the state-of-the-art recommendation model which uses review for rating prediction. User and item representation are concatenated and used as input to a Factorization Machine (FM)\cite{rendle2010factorization} for overall rating prediction.

    \item \textbf{Multi-criteria RS by Nassar et al. } \cite{nassar2020novel}. This is a deep learning based multi-criteria recommendation system. This model predicts the overall criteria ratings in two steps. In the first step, unknown k multi-criteria ratings are predicted with DNN where user id and item id are used as features. In the next and final step, the DNN model is used again to learn the relationship between multi-criteria ratings and overall ratings.
\end{enumerate}

\subsection{Experimental Settings}
First, we randomly split the data into training, validation, and test set with the ratio 80:10:10. We implement all the model with PyTorch: 1.12.1, GPU: NVIDIA Tesla P100, cuda: 11.2, and RAM: 24GB on google colab. All models are optimized with Adam optimizer. Performances are evaluated with MSE, MAE, precision, recall, and F score. For ANR, we use word2vec\cite{mikolov2013distributed} word embedding with a dimension of 300d. All parameters are kept the same as specified in the paper. For ANR and our model, we use MSELoss and train for 25 epochs with a batch size of 128. 

We implemented DeepCoNN with GloVe word embedding model \cite{pennington2014glove} for 100d and keep the other parameters as specified in the paper. Both ANR and DeepCoNN models are implemented with open-source code. For a deep multi-criteria recommendation system, code is not open sourced and we write the code and set the parameters as specified in the paper. It is important to note that, they have trained their model on small datasets crawled from the tripavisor website. They trained their model with only 72119, 1850, and 81085 users, number of items, and ratings respectively. On the other hand, we train the model with a larger dataset, with the number of users, number of items, and ratings 299855, 9763, and 577854 respectively as shown in Table II.

\begin{table*}
\centering
\caption{Performance comparison with baseline}
\begin{tabular}{c c c c c c }
\hline
Model & MSE & MAE & Precision & Recall & F1\\
\hline
\textbf{MRRRec}& \textbf{0.44043} & \textbf{0.30011} & \textbf{0.92477} &\textbf{0.95104	} & \textbf{0.93772}\\
ANR & 0.51769&0.48014	&0.89853	&0.93405	&0.91595\\
DeepCoNN & 0.94175 & 0.80059 & 0.73913 & 1.00 & 0.8500\\
Nour Nassar et. al. \cite{nassar2020novel}&0.44952	&0.30114	&0.92442	&0.94633	&0.93525\\
\hline

\end{tabular}
\end{table*}

\begin{table*}
\centering
\caption{Component Based Performance}
\begin{tabular}{c c c c c c }
\hline
Model & MSE & MAE & Precision & Recall & F1\\
\hline
Only review &0.51769&0.48014&0.89853&0.93405&0.91595\\
Only rating & 0.44638&0.31918&0.92219&0.94036 &0.93119\\
MRRRec  & 0.44043	&0.30011	&0.92477	&0.95104	&0.93772\\
\hline
\end{tabular}
\end{table*}

For a different component of our model, we use different parameter settings. The following is a set of parameters for optimal performance.
\begin{enumerate}
    \item Explicit rating prediction: we experiment the DNN model with several combination of hidden layers and different embedding dimensions of user and item ids. We randomly initialize the DNN parameters using a normal distribution with a mean of 0 and a standard deviation of 0.5. For each user and item ID, we use vector sizes from 16 to 256. For layer selection, we train the model with six different sets of layers and obtain best performance with the following combination [256$\,\to\,$128$\,\to\,$64].
    \item Implicit criteria rating prediction: since for implicit criteria rating prediction, we use components from ANR, we keep the parameters similar. However, we conducted experiments with several sets of implicit aspects $K$. We keep the size of J fixed and change the size of K and train for different sizes of K. We change the size of K because, K is a parameter that controls how many aspects we will extract from the review and by changing this parameter we can extract a different number of aspect ratings.
    \item Overall rating prediction: 
    The input dimension of the overall rating prediction model is the sum of the implicit and explicit criteria ratings.
\end{enumerate}

Performance is evaluated using MSE (Mean Squared Error), MAE (Mean Absolute Error), precision, recall and F1 score for both baselines and our proposed model. Mathematically MSE and MAE are defined as 

\begin{eqnarray}
MSE=\sqrt{\frac{1}{N} \sum_{u,i}^{N} (\hat{R} -R)^2 }
\end{eqnarray}

\begin{eqnarray}
MAE=\frac{1}{N} \sum_{u,i}^{N} |\hat{R} -R|
\end{eqnarray}
where $N$ is the total number of ratings. $\hat{R}$ and $R$ are the predicted and known overall rating respectively. 

To compute the accuracy measure, we transform our predicted ratings to 0 and 1 based on a threshold value of 3.5. The idea behind this transformation is that we decide whether to recommend an item or not based on the predicted rating. If the rating is higher than 3.5, the model makes a recommendation otherwise not. Considering these transformations, we define the problem as a binary classification problem and calculate the precision, recall, and F1 score. Precision is defined as

\begin{eqnarray}
Precision=\frac{TP}{TP+FP}
\end{eqnarray}
where $TP$ refers to true positive and $FP$ refers to false positive. For a good classifier, precision is 1.
Recall is also called the measurement of sensitivity or true positive rate.

\begin{eqnarray}
Recall=\frac{TP}{TP+FN}
\end{eqnarray}
where $FN$ is the false negative. For an ideal classifier, recall should be 1.

The F1 score is the harmonic mean of precision and recall. The F1-score measure is the metric that takes both precision and recall into account and is defined as
\begin{eqnarray}
F1 =2* \frac{Precision * Recall}{Precision + Recall}
\end{eqnarray}
F1-score is 1 only when both precision and recall are 1.

\begin{table*}[t]
\centering
\caption{Results for different Word Embedding Methods}
\begin{tabular}{c c c c c c}
\hline
Word embedding methods & MSE &MAE &Precision&Recall&F1\\
\hline
Word2Vec &0.44043	&0.30011	&0.92477	&0.95104	&0.93772\\

Glove &0.48794 &0.41067&0.92718&0.92639&0.92678\\
\hline

\end{tabular}

\end{table*}

\begin{table*}[t]
\centering
\caption{Results for different Hidden layers}
\begin{tabular}{c c c c c c}
\hline
Hidden Layers & MSE &MAE &Precision&Recall&F1\\
\hline
[256] &0.46256 &0.33712&0.923&0.94053&0.92462\\
\hline
[256, 128] & 0.45677 & 0.30746 & 0.92503 & 0.94165 & 0.93326\\
\hline
\textbf{[256, 128, 64]} & \textbf{0.44043} & \textbf{0.30011} & \textbf{0.92477} & \textbf{0.95104} & \textbf{0.93772}\\
\hline
[256, 128, 64, 32] & 0.45875 & 0.30456 & 0.92762 & 0.94253&0.93502\\
\hline
[256, 128,64, 32, 16] & 0.44441 & 0.31225 & 0.92611 & 0.94131 & 0.93365\\
\hline
[256, 128,64, 32, 16,8] & 0.43916& 0.32779& 0.92546&0.93951&0.93243\\
\hline
\end{tabular}

\end{table*}

\begin{table*}[t]
\centering
\caption{Performance of the model with different embedding sizes of ids}
\begin{tabular}{c c c c c c}
\hline
Id embedding size & MSE &MAE &Precision&Recall&F1\\
\hline
\textbf{256} &\textbf{0.44043} & \textbf{0.30011} & \textbf{0.92477} &\textbf{0.95104} & \textbf{0.93772}\\

128 &0.4476 &0.31773&0.92485&0.9389&0.93183\\

64 &0.45632 &0.34575&0.92781&0.9369&0.93233\\

32 &0.45102 &0.30936&0.92342&0.94562&0.93439\\

16 & 0.45826 & 0.32014 & 0.92636 & 0.94511 & 0.93564\\
\hline

\end{tabular}
\end{table*}

\begin{table*}[t]
\centering
\caption{Performance with different sizes of implicit criteria ratings}
\begin{tabular}{c c c c c c}
\hline
Size of K & MSE &MAE &Precision&Recall&F1\\
\hline
7 &0.45294 & 0.3104 &0.91681 &0.94314 & 0.93462\\

6 &0.45757 &0.32387&0.92313&0.94338&0.93314\\

\textbf{5} & \textbf{0.44043} & \textbf{0.30011} & \textbf{0.92477} &\textbf{0.95104} & \textbf{0.93772}\\

4 &0.45218 &0.31275 &0.92289&0.94551&0.93407\\

3 & 0.44649 & 0.29632 & 0.92956 & 0.93731 & 0.93342\\

2 & 0.45253	&0.29302	&0.92675	&0.94155	&0.93409\\
\hline

\end{tabular}

\end{table*}

\section{Results}
In this section, we present the results of different experiments and the analysis of the model.

\subsection{Results}
The final result of our proposed MRRRec comparing with the 3 baseline models is shown in Table IV. It is observed that our model outperforms all the reported baselines in all accuracy measures, MSE, MAE, precision, recall and F1 score. Our model achieves an average of 19\% and 23.3\% lower MSE and MAE respectively, and 7\%, 1.0\%, and 3.8\% higher precision, recall and F1 score respectively. It is observed that between ANR (aspect-based model) and Nour Nassar et. al. \cite{nassar2020multi} (explicit criteria rating based model), multi-criteria rating based model performs better. This can be because users express their opinions on a specific topic through explicit criteria ratings and the explicitly specified rating reflects the users' preferences more accurately than the implicitly inferred aspect rating from the review. It could also be the inability to accurately estimate the implicit criteria ratings by the existing models. Compared with aspect based model, ANR, our model achieves 7.7\%, and 18\% lower MSE and MAE respectively, and 2.6\%, 1.7\%, and 2.2\% higher precision, recall, and F1 respectively. Similarly, compared to DeepCoNN (review-based method), MRRRec achieves significantly better performance in terms of all accuracy measures. This indicates that the addition of a multi-criteria rating is important for overall rating prediction. 

\subsection{Ablation Study}
We perform an ablation study to analyze the effect of different components of the model. Table V shows the performance comparison of different network structures.
\subsubsection{Only implicit criteria}
It is observed that when we use only review data, both MSE and MAE scores are higher than the case when we use both review and criteria ratings. This review-only model achieves 7.7\% and 18.0\% higher MSE and MAE scores respectively and 2.6\%, 1.7\%, and 2.2\% lower precision, recall, and F1 score respectively.

\subsubsection{Only explicit criteria}
Similar to only implicit criteria, when we use only explicit ratings, the model is unable to capture a richer set of hidden factors from user reviews for rating prediction. With only explicit criteria ratings, the model achieves 0.6\% and 2\% higher MSE and MAE respectively, and 0.3\%, 0.5\%, and 1.0\% lower precision, recall, and F1 score respectively compared to our model. 

We also conduct experiments with two different embedding methods. The performance of Word2Vec and GloVe word embedding is reported in Table VI. It is observed that our model performs better for Word2Vec word embedding with 4.7\% and 11\% lower MSE and MAE respectively, and 0.1\%, 3\%, and 1\% higher precision, recall, and F1 score respectively. 

\subsection{Parameter Sensitivity Analysis}
We analyze the effectiveness of different parameters used in the model. Table VII shows the performance of different layer choices. It is observed that if the model is too complex, with too many hidden layers,  performance goes down. Similarly, if the model is too simple, with only a few layers, performance is also low. If we compromise the simplicity and complexity, the model performs well. We conduct experiments with different combinations of hidden layers and achieve the best performance with a hidden layer combination of [256$\,\to\,$128$\,\to\,$64]. 

We also analyze the performance of the model with a different dimension of user and item embeddings (explicit criteria rating prediction component), shown in Table VIII. It is observed that when we represent IDs with very low dimensions, the model performs poorly. As the size increases, the performance also increases slowly. We achieve the best performance with an embedding size of 256 for both user and item embeddings.

We further analyze the performance of the model with different sizes of implicit criteria. Since implicit criteria ratings are not directly provided by users and we estimate them from review, we do not know how many implicit criteria are available in the review text. Therefore, with fixed explicit criteria, we run the model for various sizes of aspects K. The effect of K is reported in Table IX. We show that the higher the value of K, the higher the value of MSE, and MAE, and the lower the value of precision, recall, and F1. We achieve the best result with $K=5$ which is equal to the number of explicit criteria.

\subsection{Discussion}
In this section, we discuss the success and failure of our model. From the obtained result, it is obvious that by integrating implicit criteria ratings into explicit criteria ratings, the model has achieved a significant improvement in performance. The existing multi-criteria rating-based recommendation system uses only explicit criteria ratings for overall rating prediction. On the other hand, an aspect-based recommendation system uses only review for overall rating prediction. Explicit criteria rating-based recommendation system is unable to capture the latent relationship between user-item pairs and the overall ratings. This is because when we consider only explicit criteria ratings, we ignore the users' latent opinions expressed through review. Similarly, when we consider only reviews, although they capture the latent preferences from reviews, they ignore users' explicit preferences on given criteria. Our model integrates both implicit and explicit criteria ratings and predicts the overall ratings.

Although we have achieved great performance improvement with our proposed model, our model is unable to extract the given explicit criteria ratings from the review. Currently, our model is incapable of specifying which implicit criteria rating is corresponding to which explicit criteria. This is because we have not trained this part of our model (left side of Figure II) with explicit criteria ratings. In the future, we will be working to map the aspect ratings to explicit criteria ratings, which would allow us to predict overall ratings more accurately. Another improvement could be to experiment with our model with several datasets. Currently to the best of our knowledge, TripAdvisor is the only publicly available dataset that provides both reviews and criteria ratings. In the future, we will focus on searching for other datasets that come with reviews and criteria ratings or creating our own dataset that can potentially benefit the research community.

We tried to evaluate our model with the most recent state-of-the-art model, however, the most recent aspect-based paper do not share the code\cite{li2021aspect} and some provided codes but without proper instruction or with missing files\cite{wu2019arp}.

\section{Conclusion}
In this work, we propose a Multi-criteria Rating and Review based Recommendation System (MRRRec). The model combines both explicit and implicit criteria ratings to predict overall ratings. The model consists of three components, namely, implicit criteria rating prediction from review texts, unknown explicit criteria rating prediction from known criteria ratings, and overall rating prediction. Implicit criteria ratings are predicted from textual reviews by aspect representation learning and aspect importance estimation. For explicit criteria rating prediction, we use a deep neural network. User IDs and item IDs are used as features and fed into the network. And finally, for overall rating prediction, we use another deep neural network. Outputs from both implicit and explicit criteria rating models are concatenated and fed as input to the final deep neural network. We show that our proposed MRRRec model outperforms three state-of-the-art models, including ANR, DeepCoNN, and Multi-criteria recommendation system in terms of MSE, MAE, precision, recall, and F1 scores. Proposed MRRRec achieves an average of 19\% and 23\% lower MSE and MAE, and 7\%, 1\%, and 3.8\% higher precision, recall, and F1 respectively. 

In the future, we plan to utilize the state-of-the-art NLP (Natural Language Processing) techniques such as BERT, and NER for aspect extraction from textual reviews as well the overall rating prediction. Furthermore, we plan to consider different Big Data Predictive Analytics frameworks \cite{cuzzocrea2009olap,souza2020innovative,cuzzocrea2013analytics} to enhance the data quality.

\section{Acknowlegements}
This work is partially sponsored by Natural Science and Engineering Research Council of Canada (grant 2020-04760)
The authors would like to thank our reviewers for their tremendous efforts on review and the constructive feedback. 

\balance
\bibliographystyle{IEEEtran}

\bibliography{library}

\vspace{12pt}
\color{red}

\end{document}